\documentclass[amssymb,aps,twocolumn,floats,showpacs]{revtex4-1}
\usepackage{color,graphicx,pstricks,float}
\usepackage{natbib}

\usepackage{fancybox}
\usepackage{epsfig}
\usepackage{graphicx}
\usepackage{tabularx}
\usepackage{inputenc}
\usepackage{amsmath}
\usepackage{amssymb}
\usepackage{gensymb}
\usepackage{array}

\begin{document}

%\title{Magnetic anisotropy prohibits Skyrmions in spin-orbit coupled 2D itinerant magnets} 
\title{Skyrmions and magnetic bubbles in spin-orbit coupled metallic magnets}

\author{Deepti Rana$^1$, Soumyaranjan Dash$^1$, Monika Bhakar$^1$, Rajeshwari Roy Chowdhury$^2$, Ravi Prakash Singh$^2$}
\author {Sanjeev Kumar$^1$}
\email {sanjeev@iisermohali.ac.in}
 \author {Goutam Sheet$^{1}$}
 \email {goutam@iisermohali.ac.in} 
\email{ Also associated with S N Bose National Center for Basics Sciences, Kolkata,  Salt Lake, JD Block, Sector III, Bidhannagar, Kolkata, West Bengal 700106 as an adjunct faculty.}
\affiliation {$^1$ Department of Physical Sciences,
Indian Institute of Science Education and Research (IISER) Mohali, Sector 81, S.A.S. Nagar, Manauli PO 140306, India \\
}

%\affiliation{$^2$ Indian Association for the Cultivation Of Science, Jadavpur, Kolkata-700032, India}

\affiliation{$^2$Department of Physics, Indian Institute of Science Education and Research (IISER) Bhopal, PO 462066, India}

%\affiliation{$^4$S. N. Bose National Centre for Basic Sciences, JD Block, Sector III, Salt Lake, Kolkata, PO 700106, India}

\begin{abstract}

%\textbf{Realization and manipulation of novel magnetic textures in metallic systems are important for their application in next-generation spintronics. Motivated by the observation of itinerant ferromagnetism in the 2D ferromagnets Fe$_n$GeTe$_2$ ($n \geq3$), we have theoretically modeled a 2D triangular lattice with spin-orbit coupling and itinerant magnetism and found a ground state with filamentary magnetic domain walls. Our magnetic force microscopy experiments on Fe$_3$GeTe$_2$ revealed strikingly similar domain walls. While the initial model predicted the formation of skyrmions under applied magnetic fields, the filaments on Fe$_3$GeTe$_2$ were experimentally seen to break into large size magnetic bubbles with magnetic field. This effect was theoretically reproduced by incorporating the role of magnetic anisotropy in the model. From our analysis of the theoretical model and the experimental data, we demonstrate how different types of topological magnetic structures can be stabilized on demand in real systems.}

{Motivated by the observation of Skyrmion-like magnetic textures in 2D itinerant ferromagnets Fe$_n$GeTe$_2$ ($n \geq3$), we develop a microscopic model combining itinerant magnetism and spin-orbit coupling on a triangular lattice. The ground state of the model in the absence of magnetic field consists of filamentary magnetic domain walls revealing a striking similarity with our magnetic force microscopy experiments on Fe$_3$GeTe$_2$. In the presence of magnetic field, these filaments were found to break into large size magnetic bubbles in our experiments. We identify uniaxial magnetic anisotropy as an important parameter in the model that interpolates between magnetic Skyrmions and ferromagnetic bubbles. Consequently, our work uncovers new topological magnetic textures that merge properties of Skyrmions and ferromagnetic bubbles. 
}

\end{abstract}
\date{\today}

\maketitle

\noindent

The topologically stable magnetic structures, such as the Skyrmions, are considered the building blocks of next-generation data storage and processing devices \cite{Fert, Laurita, Weisen, Fert1, Naga, Gobel, Karube, Bog}.
Consequently, identifying suitable magnetic materials that host such unusual magnetization textures has become a rapidly emerging area of research \cite{Dupe, Pollard, Romming, Yu, Yu1, Zhao, Meyer, Tono, Karube1, Hirschberger,Jin, Karube3}. One of the key requirements for applications is the manipulation of such textures via ultra-low electrical currents, and therefore it is desirable to have such magnetic textures realized in metallic magnets \cite{Song, Sampaio, Romming, Yu1,Ding,Han,Zhang1,Back}. The appearance of isolated Skyrmions as well as Skyrmion lattices has been reported in thin films of a variety of chiral metallic magnets \cite{Stishov, Nayak, Yu, Zhao, Pfleiderer, Jena, Meyer, Hsu, Tono, Huang, Yu2, Naga1}.
While a common understanding of magnetic Skyrmions in metals relies heavily on either a suitable free-energy functional in a continuum or classical spin models on lattices \cite{Rossler, Bog, Chen, Mohanta, Yi, Zang, Iwasaki}, more recently the importance of microscopic Hamiltonian based understanding of Skyrmion formation has been recognized \cite{Kathyat2020,Kathyat2021,Hayami,Bezvershenko}. The latter approach allows for a finer distinction between ferromagnetic bubbles and Skyrmions, thereby providing a tunable control on the stability of different topological textures.

Motivated by the recent observation of metallic ferromagnetism in the van der Waals (vdW) magnets Fe$_n$GeTe$_2$ with $n = 3,4,5$ \cite{Zhuang,Seo,May}, \textcolor{black}{we extend and investigate a recently proposed microscopic model for the broad class of vdW magnets. The model combines the effects of itinerant magnetism, spin-orbit coupling and uniaxial anisotropy \cite{Kathyat2020}.} Starting from the microscopic electronic Hamiltonian on a triangular lattice, we derive an effective spin model. We explicitly demonstrate via large-scale Monte Carlo (MC) simulations, that the ground state of the model consists of filamentary domain wall structures which show remarkable agreement with the magnetic force microscopy (MFM) images on \textcolor{black}{high-quality single crystals of } Fe$_3$GeTe$_2$. However, the MFM images in the presence of an external magnetic field deviate from the in-field calculations within the model. We show that the inclusion of a uniaxial anisotropy term in the Hamiltonian allows for a consistent description of the results at zero and finite magnetic fields. Furthermore, we identify the easy-axis anisotropy as an important tuning parameter for the relative stability of different types of topological structures in spin-orbit coupled itinerant magnets.

\noindent
In metallic magnets that consist of large magnetic moments and SOC, a generic starting model is the ferromagnetic Kondo lattice model (FKLM) in the presence of a Rashba term. We consider the Hamiltonian on a triangular lattice as,
\begin{eqnarray} \label{Ham}
H & = & - t \sum_{i,\gamma,\sigma} (c^\dagger_{i, \sigma} c^{}_{i+\gamma, \sigma} + \text{H.c.}) - J_\text{H} \sum_{i} {\bf S}_i \cdot {\bf s}_i \nonumber \\ & &
- {\textrm i}\lambda \sum_{i,\gamma, \sigma \sigma'} c_{i\sigma}^{\dagger} [\pmb{\tau}\cdot(\hat{\pmb{\gamma}} \times \hat{\bf{z}})]_{\sigma \sigma'} c_{j\sigma'} - h_z \sum_{i} S_i^z.
\end{eqnarray}

\noindent
The annihilation (creation) operators, $c_{i\sigma}$ ($c_{i\sigma}^\dagger$), satisfy the usual Fermion algebra. $J_{\rm H}$ ($\lambda$) denotes the strength of Kondo (Rashba) coupling, $t$ is the nearest neighbor hopping parameter on triangular lattice. $\pmb{\tau}$ is a vector operator with the three Pauli matrices as components. $\bf{s}_i$(${\bf S}_i$) denotes the electronic spin operator (localized classical spin) at site $i$. Assuming the lattice constant to be unity, $\hat{\pmb{\gamma}} \in \{ {\bf a}_1, {\bf a}_2, {\bf a}_3 \}$ are the primitive vectors of the triangular Bravais lattice with ${\bf a}_1$=(1,0), ${\bf a}_2$=(1/2,$\sqrt{3}/2$) and ${\bf a}_3$=(-1/2,$\sqrt{3}/2$). The last term in Eq. (\ref{Ham}) represents the Zeeman coupling of local moments to an external magnetic field of strength $h_z$. $t = 1$ sets the basic energy unit in the model. \textcolor{black}{Presence of Skyrmion Hall effect in the vdW magnets indicates that the strong coupling limit is more relevant as compared to the weak-coupling RKKY limit \cite{Wu,Akosa,You,Wang}. This leads to the Rashba double-exchange(RDE) Hamiltonian on a triangular lattice} \cite{Kathyat2020},

\begin{figure*}
\includegraphics[width=.75 \textwidth,angle=0,clip=true]{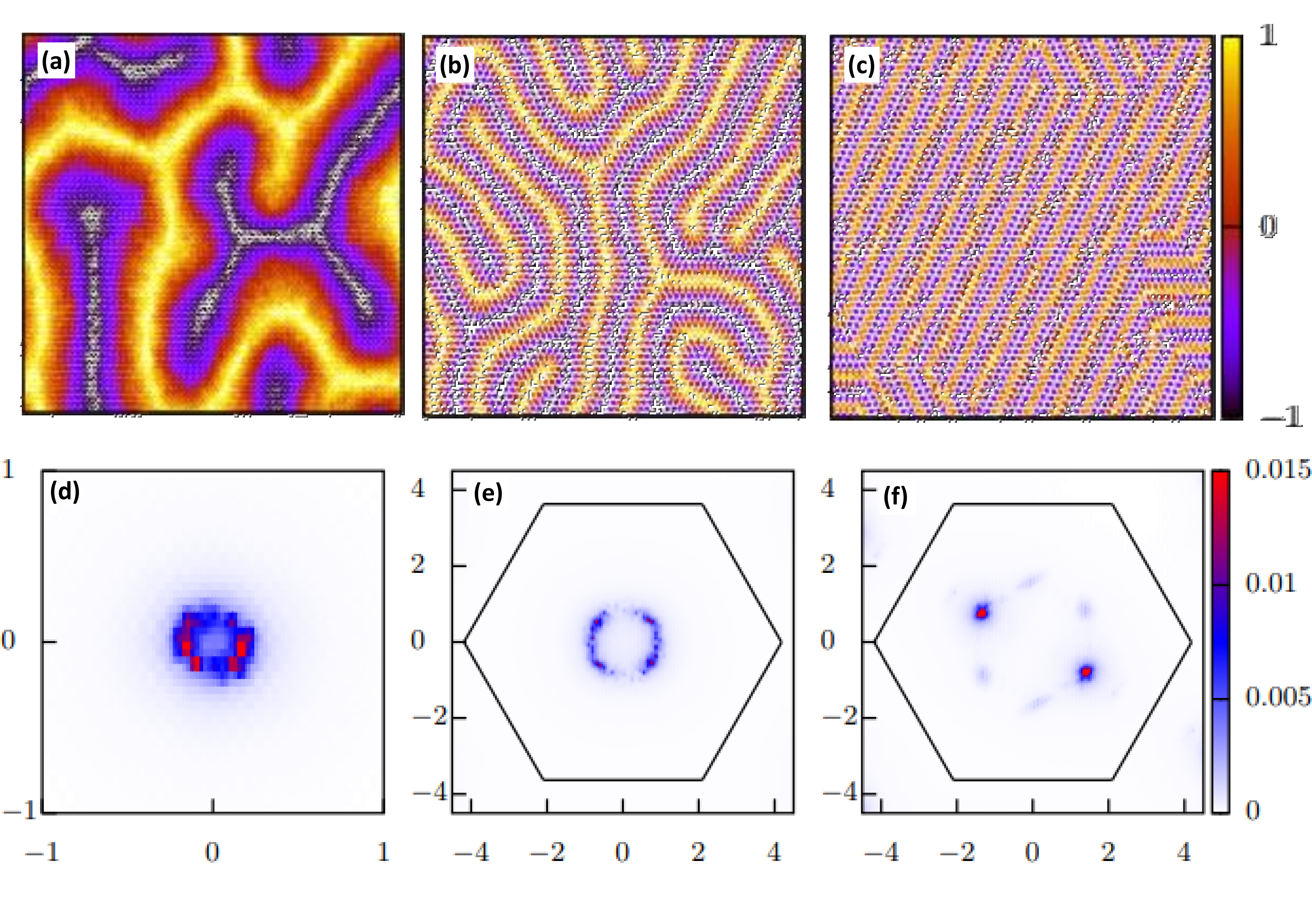}
\caption{(Color online) Real-space view of spin configurations at low temperature ($T/t = 0.01$) for (a) $\lambda/t = 0.1$, (b) $\lambda/t = 0.5$ and (c) $\lambda/t = 1$. The color bar corresponds to the z-component, and the arrows indicate the planar components of the spins. (d)-(f) The corresponding spin structure factors for the three values of the SOC strength $\lambda/t$. 
}
\label{fig1}
\end{figure*}

\begin{eqnarray}
H_{\rm RDE} &=& \sum_{\langle ij \rangle, \gamma} [g^{\gamma}_{ij} d^{\dagger}_{i} d^{}_{j} + {\rm H.c.}] - h_z \sum_i S^z_i,
\label{Ham-RDE}
\end{eqnarray}
\noindent
where, $d^{}_{i} (d^{\dagger}_{i})$  annihilates (creates) an electron at site ${i}$ with spin parallel to the localized spin.  Site $j = i + \pmb{\gamma}$ is the nn of site $i$ along one of the three symmetry directions on the triangular lattice. The projected hopping $g^{\gamma}_{ij}$ depend on the orientations of the local moments ${\bf S}_i$ and ${\bf S}_j$. The tight-binding, $t^{\gamma}_{ij}$ and Rashba, $\lambda^{\gamma}_{{ij}}$, contributions to $g^{\gamma}_{ij} = t^{\gamma}_{ij} + \lambda^{\gamma}_{ij}$ are given by \cite{Kathyat2020},

\begin{eqnarray}
t^{\gamma}_{ij} & = & -t \big[\cos(\frac{\theta_i}{2}) \cos(\frac{\theta_j}{2}) 
+ \sin(\frac{\theta_i}{2})  \sin(\frac{\theta_j}{2})e^{-\textrm{i} (\phi_i-\phi_j)} \big],
\nonumber \\
 & &\lambda_{{ij}}^{{\bf a}_1}=\lambda_{{ij}}^x, \ \ \ \ \ 
 \lambda_{{ij}}^{{\bf a}_{2/3}} = \pm \frac{1}{2}\lambda_{{ij}}^x + \frac{\sqrt{3}}{2}\lambda_{{ij}}^y, \nonumber \\
\lambda_{{ij}}^x & = & \lambda \big[\sin(\frac {\theta_i}{2})  \cos(\frac {\theta_j}{2})e^{-\textrm{i} \phi_i} - \cos(\frac {\theta_i}{2})  \sin(\frac {\theta_j}{2})e^{\textrm{i} \phi_j}\big],
\nonumber \\ 
\lambda_{{ij}}^y & = & \textrm{i} \lambda \big[\sin(\frac {\theta_i}{2})  \cos(\frac {\theta_j}{2})e^{-\textrm{i} \phi_i} + \cos(\frac {\theta_i}{2})  \sin(\frac {\theta_j}{2})e^{\textrm{i} \phi_j}\big]
\end{eqnarray}

\noindent
where $\theta_i$ ($\phi_i$) is the polar (azimuthal) angle for localized moment ${\bf S}_i$. 

\begin{figure*}
\includegraphics[width=.75 \textwidth,angle=0,clip=true]{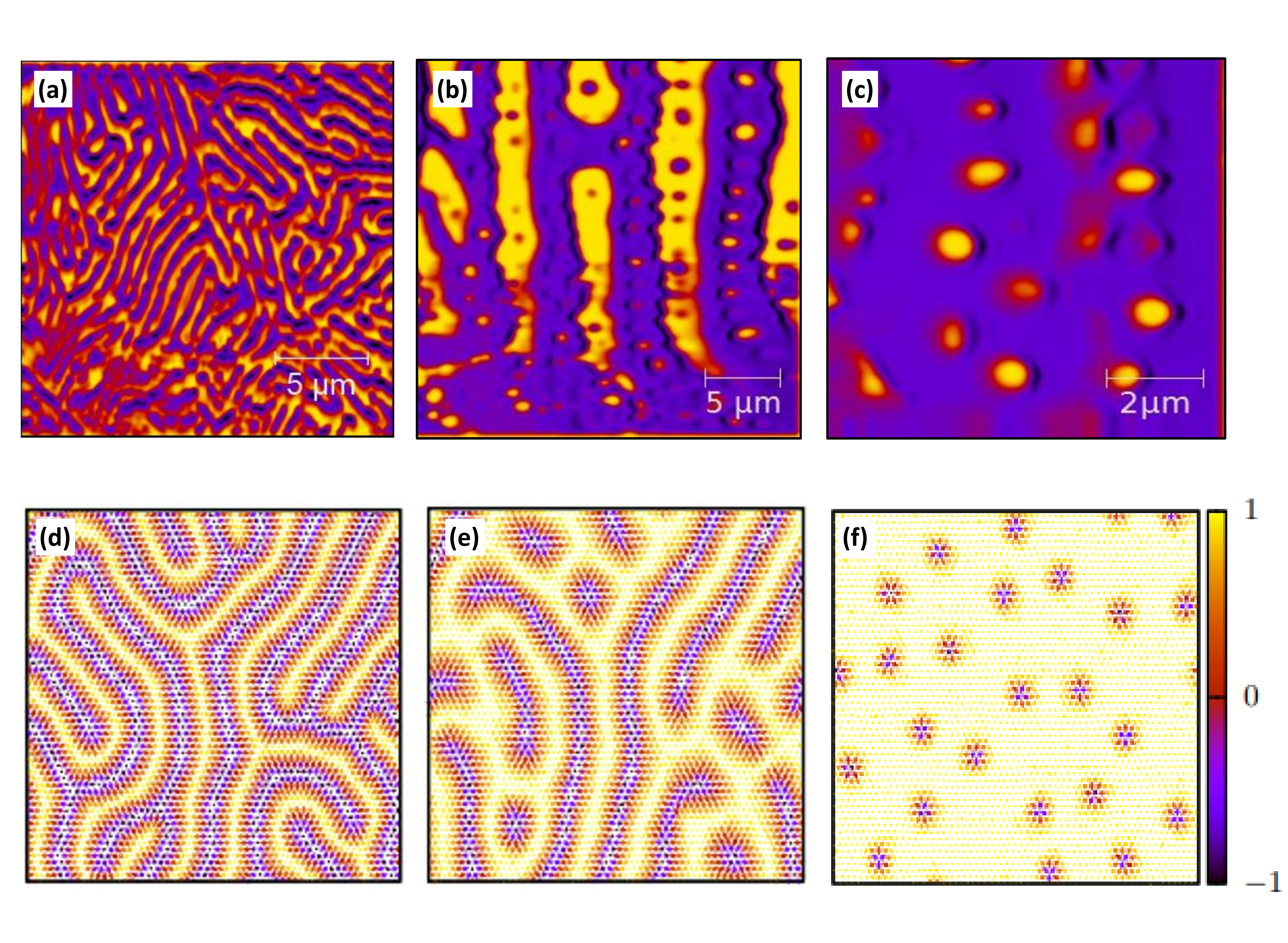}
\caption{MFM dual pass phase images taken on a cleaved single crystal of Fe$_3$GeTe$_2$ where (a) was taken at 1.6 K in the zero field cooled (ZFC) state. The filamentary domains can be seen clearly. (b)-(c) were imaged in the field-cooled state in the presence of the field of the MFM cantilever.  (Color online) Representative spin configurations at $T/t = 0.01$ and $\lambda/t = 0.5$ for (d) $h_z = 0.2$, (e) $h_z = 0.44$ and (f) $h_z = 0.84$. The color bar corresponds to the $z$-component, and the arrows indicate the planar components of the spins.
}
\label{fig2}
\end{figure*}

\noindent 

The Hamiltonian Eq. (\ref{Ham-RDE}) describes a modified tight-binding model where the hopping integrals are dependent on the configuration of classical spins. Therefore, the energy of the system depends on the classical spin configurations. This dependence can be formally written as an effective spin Hamiltonian by following a procedure well known for double-exchange models \cite{Kumar, Kathyat2020}. For the present case of Rashba coupling on a triangular lattice, we obtain,

\begin{eqnarray}   \label{eq:ESH} 
H_{\rm eff} & = &-\sum_{\langle ij \rangle, \gamma} D^{\gamma}_{ij} f^{\gamma}_{ij} - h_z \sum_i S^z_i,  \nonumber \\
\sqrt{2} f^{\gamma}_{ij} & = &  \big[ t^2(1+{\bf S}_i \cdot {\bf S}_j) + 2t\lambda  \hat{\pmb{\gamma}'} \cdot ({\bf S}_i \times{\bf S}_j)  \nonumber  \\
& & + \lambda^2(1-{\bf S}_i \cdot {\bf S}_j+2 (\hat{\pmb{\gamma}'} \cdot {\bf S}_i)(\hat{\pmb{\gamma}'} \cdot {\bf S}_j)) \big]^{1/2}, \nonumber \\
D^{\gamma}_{ij} & = & \langle [e^{{\rm i} h^{\gamma}_{ij}} d^{\dagger}_{i} d^{}_{j} + {\textrm H.c.}] \rangle_{gs}.
\end{eqnarray} 

\noindent
In the above, $f^{\gamma}_{ij}$ ($h^{\gamma}_{ij}$) is the modulus (argument) of complex number $g^{\gamma}_{ij}$, and $\langle \hat{O} \rangle_{gs}$ denotes expectation values of operator $\hat{O}$ in the ground state. 
%Given that the form of the Hamiltonian is similar to that obtained for Rashba SOC, we expect the physics for $h_z=0$ to be similar \cite{Kathyat2020}.
Assuming constant coupling parameters has been shown to be a good approximation for studying ground state phases of $H_{\rm eff}$ \cite{Kathyat2020}, therefore we set $D^{\gamma}_{ij} \equiv D_0 = 1/\sqrt{2}$ to study $H_{\rm eff}$. We perform Monte Carlo (MC) simulations on Hamiltonian Eq. (\ref{eq:ESH}) via the standard Markov chain MC using Metropolis algorithm. The details of this computational technique can be found in the supplementary material section II \cite{SMmm}.

\begin{figure*}
\includegraphics[width=.75 \textwidth,angle=0,clip=true]{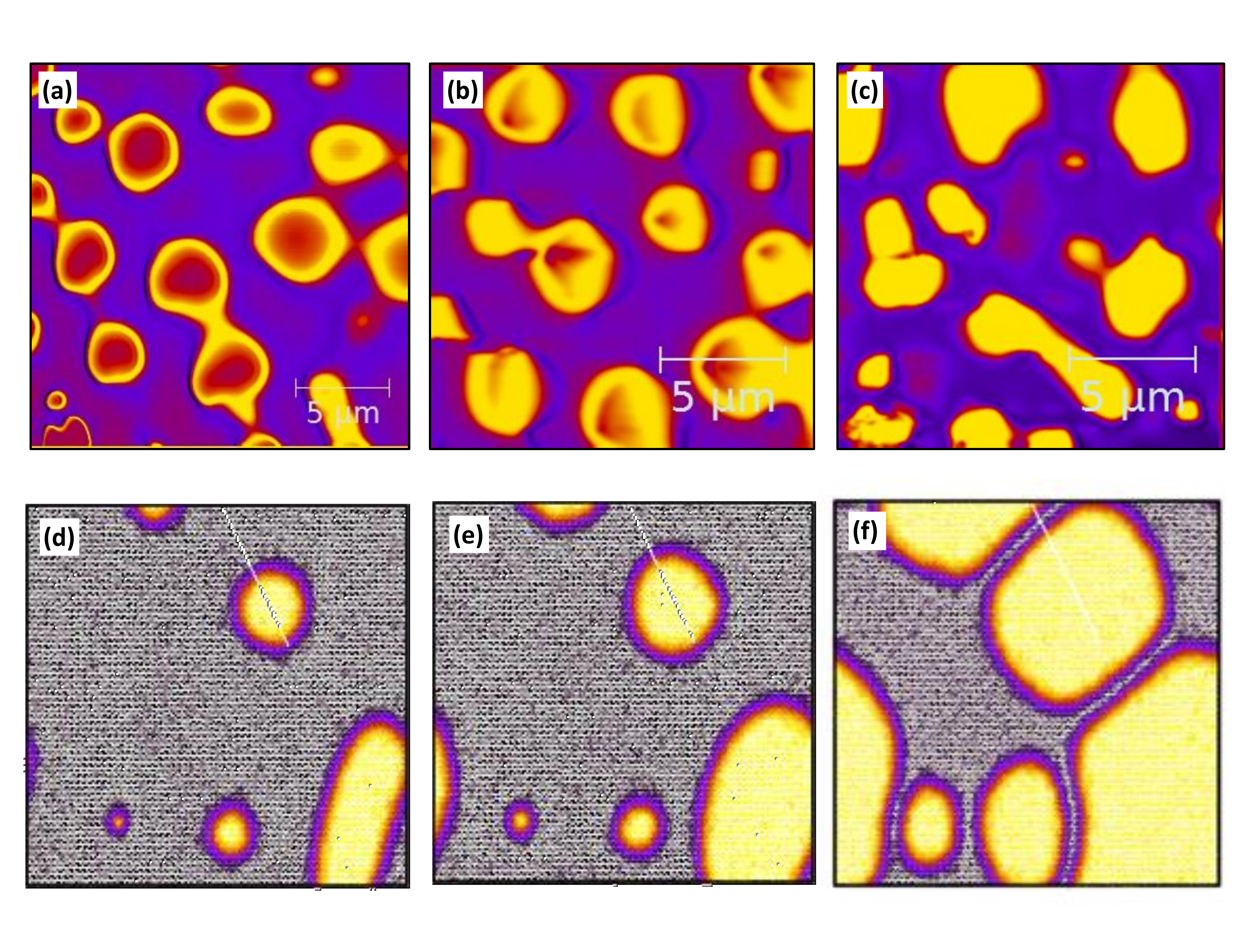}
\caption{(a)-(c) MFM dual pass images,  \textcolor{black}{recorded at different regions on the sample surface}, in the field-cooled state (FC) where an external field of 1.2 kOe was applied. The lift height was kept constant at 30 nm for all the images. (Color online) Representative spin configurations at $T/t = 0.01$, $\lambda/t = 0.1$ and $A_u = 0.08$ for (d) $h_z = 0.004$, (e) $h_z = 0.016$ and(f) $h_z = 0.028$. %The color bar corresponds to the $ z$ component, and the arrows indicate the planar components of the spins.
}
\label{fig3}
\end{figure*}

The low-temperature magnetic states, with varying strengths of SOC, are shown in Fig. \ref{fig1}. For small to intermediate values of $\lambda/t$, we find the filamentary domain structure of spins (see Fig. \ref{fig1}(a)-(b)). The thickness of these domains decreases with the increasing strength of SOC. For a sufficiently large lattice size, we find that the filaments are oriented along all possible directions, and this is more clearly reflected in the structure factor plots where a circular pattern in the $S({\bf q})$ is obtained (see Fig. \ref{fig1}(d)-(e)). This freedom of the domains to orient freely in any direction is a consequence of the anisotropic DM interaction encoded in the linear in $\lambda$ term in the $f_{ij}$ in Hamiltonian Eq. (\ref{eq:ESH}). For larger values of $\lambda/t$, we obtain a spin-spiral state with a single ordering wavevector (see Fig. \ref{fig1}(c), (f)). The origin of this state is related to the importance of the $\lambda^2$ term in the effective Hamiltonian. Note that the $\lambda^2$ term prefers Ising-like interactions of different components along different directions, leading to a classical Kitaev model with degenerate ground states. Therefore, there are three states, related by the rotational symmetry of the triangular lattice, that can be realized as ground states. In terms of the SSF, the symmetry-related states will display a SSF that is already rotated by $2\pi/3$ and $4\pi/3$ w.r.t. Fig. \ref{fig1}(f). \textcolor{black}{Note that the $\lambda^2$ term resembles a truncated dipolar interaction term. Therefore, similar magnetic domains are likely to be present in any simulation that includes dipolar interactions \cite{Birch}.}

 Now, we focus on the experimentally observed magnetic domain structures on Fe$_3$GeTe$_2$  where the characteristic features of the model, namely the existence of ferromagnetism \cite{Deiseroth}, metallicity \cite{Zhuang}, and high spin-orbit coupling \cite{Kim} exist naturally. Fe$_3$GeTe$_2$ is a van der Waals layered material exhibiting a hexagonal lattice structure with spacegroup $P6_3$/mmc. It is an itinerant ferromagnet with a Curie temperature ranging from 220-230 K in bulk \cite{Deiseroth,Chen1}. \textcolor{black}{High-quality single crystals of Fe$_3$GeTe$_2$  were synthesized by chemical vapor transport (CVT). Further discussion on the sample
growth parameters and the characterization is included in the supplemental material section I \cite{SMmm}}. We imaged the magnetic domains in Fe$_3$GeTe$_2$ by magnetic force microscopy (MFM) performed at different temperatures and under magnetic fields. At 1.6 K, under zero field cooled (ZFC) condition, we found filamentary domains with a typical domain width of around 900 nm as shown in Fig. \ref{fig2}(a). The domain structure involves stripes and interconnects with striking similarities with the computationally obtained filamentary domains as shown in Fig. \ref{fig1}(b). \textcolor{black}{Note that the domain width obtained in simulations is much smaller compared to what is observed in FGT. However, we explicitly verify that simulations on larger lattices with smaller $\lambda$ lead to wider domains \cite{SMmm}. Therefore, the mechanism of the formation of such domains is correctly captured within our approach.}
%Therefore, it appears that Fe$_3$GeTe$_2$ hosts moderate spin-orbit coupling with $\lambda/t \sim 0.5$. 
\textcolor{black}{It is important to note that, unlike stripe domains \cite{Voltan}, which consist of wide,
parallel stripes of alternating magnetic orientations, or fractal domains \cite{Kreyssig,Catalan}, which exhibit intricate and irregular patterns, filamentary domain wall structures are characterized by their
continuous nature. In certain magnetic materials, especially those with strong magnetic
anisotropy and competing interactions, the formation of filamentary domain wall structures
can be energetically favorable.
}
In order to gain a further understanding of the magnetic state of Fe$_3$GeTe$_2$, we performed MFM imaging in the field-cooled (FC) states of the crystal. The details of the MFM technique can be found in the supplementary material section II \cite{SMmm}.  We first field-cooled the crystal under the magnetic field of the MFM cantilever and then performed MFM imaging at 1.6 K. Under this condition, as shown in Fig. \ref{fig2}(b), we found that circular domains (the dark circles) have formed within the stripes with the magnetization pointing opposite to the direction of the tip-magnetization, along with rows of circular domains with the magnetization aligned in the opposite direction (the bright circles). This feature was reported earlier in Fe$_3$GeTe$_2$ and was related to high perpendicular magnetic anisotropy \cite{Nguyen,Brito}. Interestingly, at several points, under the field, the stripes have also started breaking up into fragments. To investigate the microscopic structure in further detail, we imaged a smaller area (see Fig. \ref{fig2}(c)) under the same condition where we find an assembly of bright circular domains with a background hosting the fragmented stripes. Though the circular domains have a close resemblance with typical Skyrmions, their size is relatively larger ($\sim 500\mu m$) and may be called ``magnetic bubbles". These differ from the calculated domain structures within our model. In the model, the zero-field cooled protocol was followed with the temperature reduced to $T/t = 0.01$ in the absence of a magnetic field followed by the increase of field in a step-wise manner until a fully saturated ferromagnetic state is obtained. Within the model, for intermediate values of the SOC strength, which is relevant for Fe$_3$GeTe$_2$ as noted earlier, the filamentary domains are seen to gradually break into well-defined Skyrmions upon increasing the magnetic field strength (see Fig. \ref{fig2}(d)-(f)). The results for small and high SOC strengths are discussed in the supplementary material section III \cite{SMmm}.

In search for a deeper understanding of the difference
between the experiment and the model, we studied the
effect of a higher external field.   For that, the crystal was first warmed upto 300 K and was then field cooled under an applied magnetic field of 1.2 kOe. Subsequently, the magnetic domains were imaged at zero applied field. As shown in Fig. \ref{fig3}(a), in a given area, we found that the stripes of bright domains with circular dark regions have broken into parts and formed individual circular domains with darker central region with brighter perimeters. There are regions where two such domains are seen to be connected where the breaking-up process remained incomplete. While such circular domains are in majority, flat bright domains are also seen in the same region. In another region, where the surface is slightly tilted with respect to the MFM tip, the domains were imaged from an oblique direction which revealed a \textcolor{black}{conical-like} shape of the domains (see Fig. \ref{fig3}(b)). This means, in Fig. \ref{fig3}(d), the ring like domains are merely the top-view of \textcolor{black}{such conical-shaped domains}. 
We then moved to an area where the bright flat domains are more in number, in order to investigate the distribution of their size and shape. As shown in Fig. \ref{fig3}(c), the flat domains appear to be fragments of the stripes in Fig. \ref{fig2}(a). The domains have condensed into different shapes including nearly circular and elongated ones. All these observations differed significantly from the results of the model (see Fig. \ref{fig2}(d)-(f)). 

To account for the discrepancy discussed above, we note that the model discussed so far does not include an important feature of the material, {\it i.e.}, the presence of easy-axis anisotropy. It is important to ask if the presence of easy axis anisotropy can explain the presence of large ferromagnetic bubbles in Fe$_3$GeTe$_2$ and the domain structure at higher magnetic fields. In order to investigate this, we included an additional term, $-A_u \sum_i (S^z_i)^2$, in the Hamiltonian Eq. (\ref{eq:ESH}). Positive values of $A_u$ represent an easy-axis scenario relevant to the material. \textcolor{black}{The values of $A_u$ and $h_z$ used in simulations are such that the ratio of magnetization to the saturation magnetization, and the magnetization values for field along c-axis and perpendicular to c-axis compare well with the corresponding experimental values \cite{SMmm}.}
We find that the anisotropy scale competes with the \textcolor{black}{Dzyaloshinskii–Moriya}(DM) terms and therefore, the tendency of spins to continuously tilt away from the single Skyrmion-core spin is suppressed. This leads to an expansion of the Skyrmion core and finite regions with ferromagnetic bubble character are stabilized (see Fig. \ref{fig3}(d)-(f)). These are consistent with the larger size ``magnetic bubbles" that were seen experimentally under the field of the cantilever. Further, in the model, we find that upon increasing magnetic field strength the FM regions grow in size while remaining trapped inside self-enclosing domain walls. These are similar to the ring-like domain structures seen at 1.2 kOe. At even higher magnetic fields the model generates larger size flat regions that show striking similarities with the flat regions seen experimentally. The variation of the experimental domain structure from point to point could be attributed to an inhomogeneity of the surface properties which might arise from an inhomogeneous distribution of the Fe vacancies in the system. Therefore, the results obtained in the simulations including an easy-axis anisotropy are now consistent with the evolution of the MFM images with applied magnetic field.

\noindent
%{\it Conclusion:--}

In conclusion, motivated by the demonstration of metallicity as well as ferromagnetism in the 2D materials Fe$_n$GeTe$_2$ ($n \geq 3$), we developed a model to understand the origin and the nature of topologically stable magnetic textures in a class of itinerant magnets. The model is generally applicable to spin-orbit coupled systems that display itinerant magnetism. Results of Monte-Carlo simulations on our model display remarkable agreement with our experimental data on Fe$_3$GeTe$_2$ in identifying filamentary domain walls as the ground state structure in the absence of magnetic field. In the presence of magnetic field the model predicts either isolated Skyrmions or Skyrmion lattice, depending on the strength of the SOC. Our MFM experiments, on the other hand, reveal large size magnetic bubbles with certain additional features. These magnetic structures were understood within the model when an easy-axis a nisotropy term, as relevant to the materials under discussion, was included. 
%Consequently, we have established a systematic interpolation between the concept of an ideal Skyrmion and that of a ferromagnetic bubble. 
%This leads to an important generalization of the concept of topologically stable magnetization textures by including the Skyrmionic bubble. Therefore, our results provide a basis to design new materials with distinct topologically stable magnetic structures in low-dimensional metallic ferromagnets, through the tuning of spin-orbit coupling and magnetic anisotropy.

Our combined experimental and theoretical investigation on a class of vdW itinerant magnets has allowed us to identify easy-axis anisotropy as an important parameter that can tune the nature of topological textures from Skyrmions to magnetic bubbles. This generalization of Skyrmion-like topological textures establishes a conceptually new framework for characterizing real-space imaging data of experiments on itinerant magnets.

GS acknowledges financial assistance from the Science and Engineering Research Board (SERB), Govt. of India (grant number: \textbf{CRG/2021/006395}). GS also acknowledges S N Bose National Center for Basic Sciences, Kolkata for faculty fellowship. DR and MB thank DST Department of Science and Technology (DST), Govt.
of India for INSPIRE Fellowship. We acknowledge the use of the computing facility at IISER Mohali. R. R.C acknowledges
Department of Science and Technology (DST), Govt.
of India, for financial support (Grant no.
\textbf{DST/INSPIRE/04/2018/001755}) R.P.S. acknowledges the Science and Engineering Research Board (SERB), Govt. of India, for Core Research Grant (\textbf{No. CRG/2019/001028}).

\end{document}

\title{Supplemental Material For  \\
Skyrmions and magnetic bubbles in spin-orbit coupled metallic magnets}

\author{Deepti Rana$^1$, Soumyaranjan Dash$^1$, Monika Bhakar$^1$, Rajeshwari Roy Chowdhury$^2$, Ravi Prakash Singh$^2$}
\author {Sanjeev Kumar$^1$}
\email {sanjeev@iisermohali.ac.in}
 \author {Goutam Sheet$^{1}$}
 \email {goutam@iisermohali.ac.in}
\email{ Also associated with S N Bose National Center for Basics Sciences, Kolkata,  Salt Lake, JD Block, Sector III, Bidhannagar, Kolkata, West Bengal 700106 as an adjunct faculty.}
\affiliation {$^1$ Department of Physical Sciences,
Indian Institute of Science Education and Research (IISER) Mohali, Sector 81, S.A.S. Nagar, Manauli PO 140306, India \\
}

%\affiliation{$^2$ Indian Association for the Cultivation Of Science, Jadavpur, Kolkata-700032, India}

\affiliation{$^2$Department of Physics, Indian Institute of Science Education and Research (IISER) Bhopal, PO 462066, India}

\begin{abstract}
    This supplementary material includes (i) sample details; (ii) the details of the computational and experimental techniques used; and (iii) a discussion on the evolution of the domain structures with varying spin-orbit coupling strengths in the presence of an applied magnetic field.
\end{abstract}

\maketitle

\section{Sample details:} 
High-quality single crystals of Fe$_{3}$GeTe$_{2}$, synthesized by chemical vapor transport, were used for our magnetic force microscopic (MFM) measurements. A stoichiometric mixture of the ingredients elements ( Fe (3N), Ge (3N), and Te (3N)) in the powder form was sealed in an evacuated quartz tube along with I$_{2}$ as the transport agent. The tube was kept in a two-zone furnace at a temperature gradient of 750$^\circ$C/650$^\circ$C. Plate-like single crystals were obtained after two weeks.
Figure S1(a) shows the optical microscope image of a single crystal of Fe$_{3}$GeTe$_{2}$. Figure S1(b) shows the experimental out-of-plane X-ray diffraction (XRD) results for the single crystal where sharp peaks were observed. The observed Bragg peaks can be indexed with (00$l$) peaks. Figure S1(c) shows the Laue diffraction pattern of a Fe$_{3}$GeTe$_{2}$ crystal, confirming the six-fold symmetry of the hexagonal structure and high crystallinity of the grown sample. The chemical composition of the grown crystals was confirmed from atomic percentage ratios obtained from energy-dispersive X-ray (EDX) spectroscopy measurements within the instrumental limit (Figure S1(d)). Further characterization details of the sample are reported elsewhere \cite{Chowdhury}.
\begin{figure*}
\centering
\includegraphics[width=0.75 \textwidth,angle=0,clip=true]{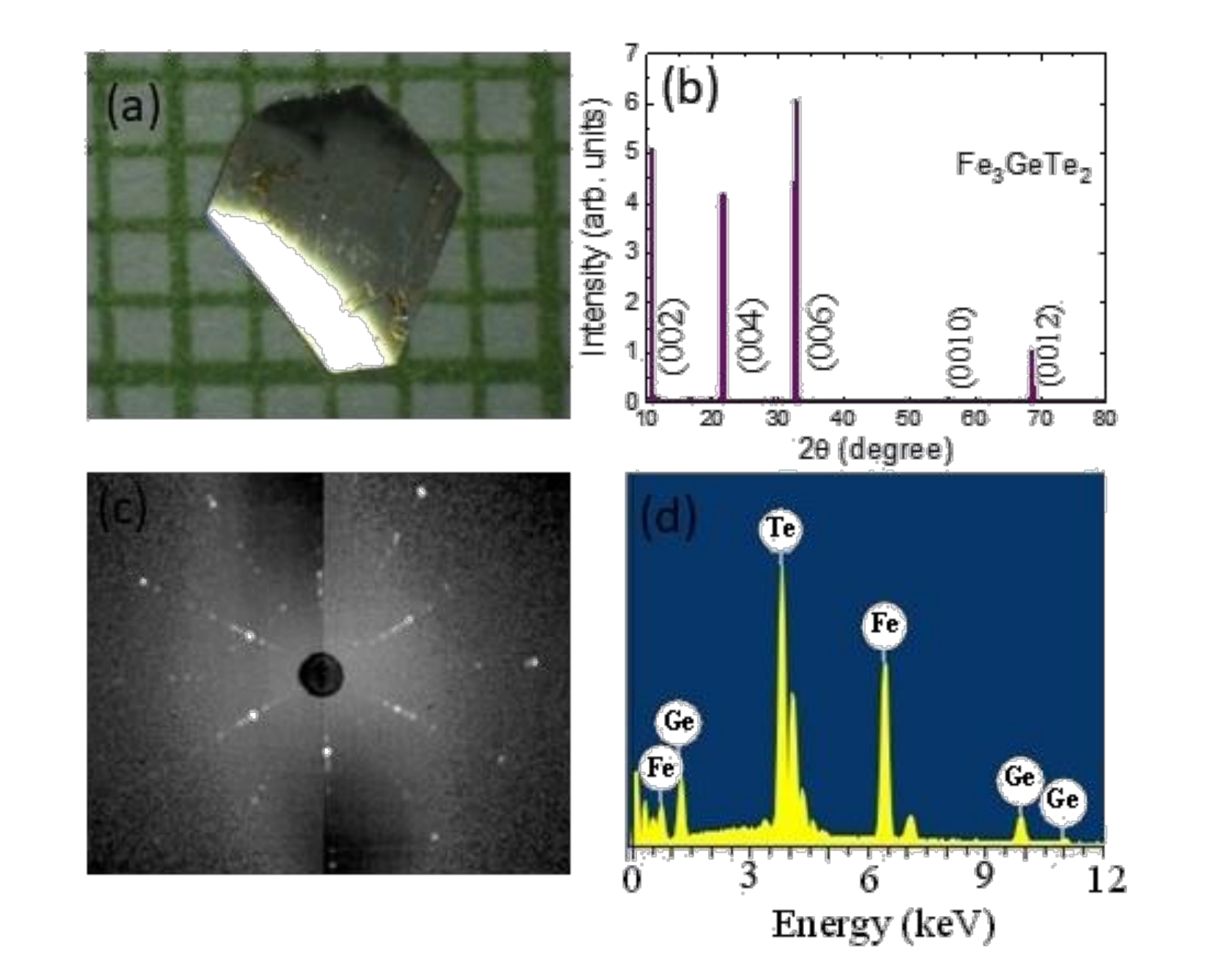}
\caption{(a) Optical micrograph of single crystal of Fe$_{3}$GeTe$_{2}$ utilized in the measurements. (b) Out-of-plane X-ray diffraction pattern of Fe$_{3}$GeTe$_{2}$ at room temperature (c)  Laue diffraction pattern obtained on Fe$_{3}$GeTe$_{2}$ single
crystal. (d) EDX spectra for Fe$_{3}$GeTe$_{2}$.}

\end{figure*}
\section{Methods}
\textbf{Magnetic force microscopy:} In magnetic force microscopy (MFM), a silicon cantilever/tip coated with a magnetic coating ( typically Co
or Al) is used along with an interferometer-based detection system to probe the local magnetic
properties of the sample by measuring the magnetic interactions between the tip and the sample. A
low-temperature compatible magnetic force microscope (Attocube LT-MFM), working down to 1.6 K, equipped with a single mode fiber-based interferometer was used for performing
the ferromagnetic domain imaging on both the ferromagnets. The phase images presented in this
work were obtained in dual-pass mode also known as lift mode. In the single pass, the cantilever
using a feedback loop maps the topography of the sample surface where the effect of van der Waal
interactions dominates. Since the magnetic forces are long-ranged as compared to the van der Waals
forces, the effect of the topographic variations can be eliminated to obtain a pure magnetic signal
by lifting the tip to a certain height above the sample. For image processing, Gwyddion software \cite{Necas}
was used.

 \textbf{Monte-Carlo simulations:} We begin the simulations at high temperature and reduce temperature in small steps, using $10^5$ MC steps for equilibration at each temperature. The magnetic field value is kept zero during this cooling process. Finally, at low but finite temperature we increase magnetic field in step-wise manner. Once again $10^5$ MC steps for equilibration are used at each magnetic field value, followed by an equal number of steps for recording averages of observables. Most importantly, the MC approach allows access to full spatial details that can be used not only to identify unusual magnetization textures but also to compare with spatially resolved experimental data as will be discussed later. Therefore, our analysis focuses on the typical spin configurations obtained at low temperatures and their evolution with applied magnetic field. We also compute the spin structure factor, $S({\bf q})$, in order to identify the presence of ordered magnetic phases. The spin structure factor is given by,
$$
S({\bf q}) = \frac{1}{N^2} \left \langle \sum_{i,j} {\bf S}_i \cdot {\bf S}_j~ \exp[-i ({\bf r}_i-{\bf r}_j)\cdot {\bf q}] \right \rangle,
$$
\noindent
where $N$ is the number of sites, and the angular bracket denotes averaging over MC configurations.
\section{Evolution with magnetic field for different SOC strengths.}
 For small SOC, the filamentary domains are wide
and the presence of a magnetic field leads to the generation of isolated large Skyrmions (see Fig. S2 (a)-(c)). The typical
size of skyrmions obtained here can be estimated from
the inverse of the radius of the circular pattern in SSF
(see Fig. S2 (d)-(e)). Since the Skyrmions do not form
any pattern, their presence is difficult to infer from the
SSF. Indeed, the SSF shown in Fig. S2 (f) indicates a ferromagnetic phase via a single peak at the $\Gamma$ point. The case of the intermediate values of the SOC strength is discussed in the manuscript. For strong SOC, the zero-field state is 3-
fold degenerate stripe state. This leads to the formation
of a densely packed hexagonal pattern of Skyrmions at finite fields (see Fig. S4 (a)-(c)). As expected, the SSF (Fig.
S4 (d)-(f)) is more useful for identifying the formation of
Skyrmions in this case as they organize into a lattice. The physics in the formation of filamentary domains in different SOC limit is similar, except for the domain width. For smaller SOC strength, we need relatively larger lattice to see the domains clearly . In order to explicitly demonstrate the validity of our claim, we have simulated $N = 240^2$ system, and the results are displayed in Fig. S5. It can be clearly seen that these filamentary domains gets more wide as SOC strength decreases. Including easy axis anisotropy in the model leads to the transition of filamentary domains to large-sized ferromagnetic bubbles as shown in Fig. S6.
The correspondence between the magnetic field values used in the experiment and the simulations is an important one, and the best way to compare the parameters is to compare the $M/M_{{\rm sat}}$ ratio at the applied field values. We find that $M/M_{{\rm sat}} \approx 0.2$ at $1.2$ kOe. For $h_z = 0.028$ (corresponding to Fig. 3c in the manuscript), we find $M/M_{{\rm sat}} = 0.15$. Furthermore, the value of anisotropy parameter $A_u = 0.08$ used in the simulations can be justified by comparing between experiment and simulations the ratio of the value of magnetization with applied field oriented along the c-axis to that oriented perpendicular to the c-axis. We find that for the values of magnetic fields mentioned above, $M_z (h || z)/M_x (h || x ) \approx 3$ in experiments \cite{Chowdhury} as well as the simulations.

\begin{figure*}
\centering
\includegraphics[width=1 \textwidth,angle=0,clip=true]{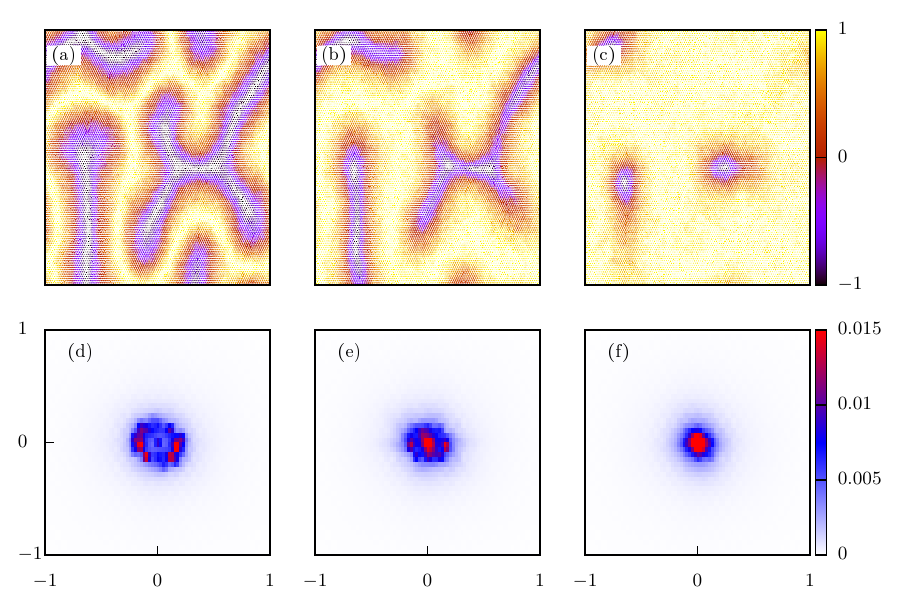}
\caption{ (Color online) Representative spin configurations at $T /t = 0.01$ and $\lambda/t = 0.1$ for (a) $h_z = 0.008$, (b) $h_z = 0.024$ and
(c) $h_z = 0.032$. The color bar corresponds to the z-component, and the arrows indicate the planar components of the spins. (d)-(f) The corresponding spin structure factor for the three values of the magnetic field $h_z$.}

\end{figure*}

\begin{figure*}
\centering
\includegraphics[width=1 \textwidth,angle=0,clip=true]{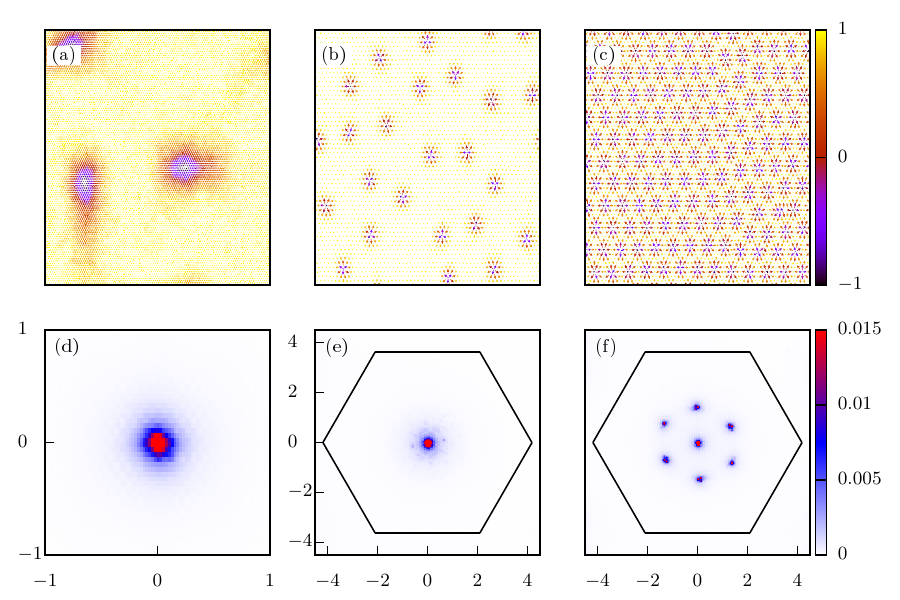}

\caption{  (Color online) Representative spin configurations at $T /t = 0.01$ for (a) $\lambda/t = 0.1$, $ h_z = 0.032$, (b) $\lambda/t = 0.5$, $h_z = 0.84$ and (c) $\lambda/t = 1.0$, $h_z = 1.40$. The color bar corresponds to the z-component, and the arrows indicate the planar components
of the spins. (d)-(f) The corresponding spin structure factor for the three values of the SOC and magnetic field $h_z$.}
\end{figure*}

\begin{figure*}
\centering
\includegraphics[width=1 \textwidth,angle=0,clip=true]{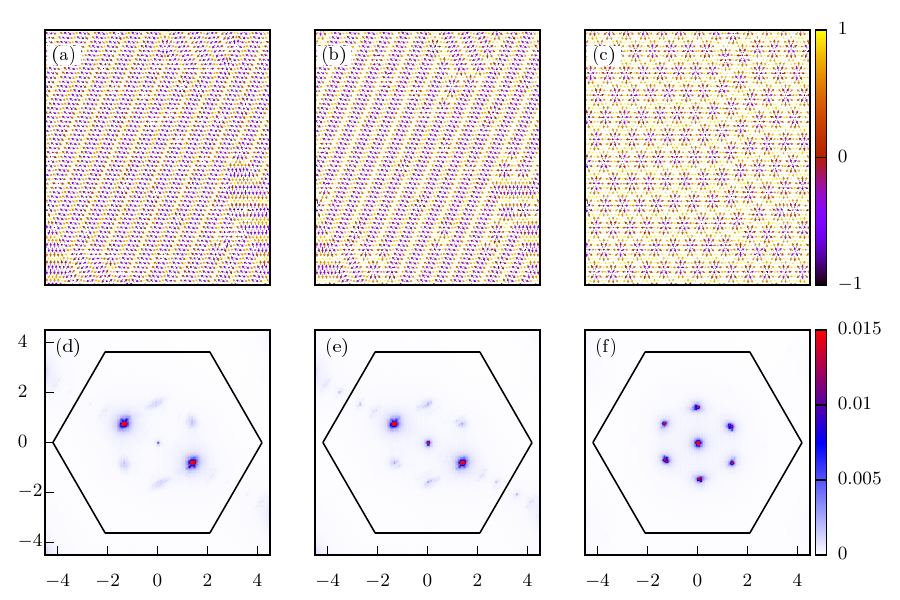}
\caption{  (Color online) Representative spin configurations at $T /t = 0.01$ and $\lambda/t = 1$ for (a) $h_z = 0.4$, (b) $h_z = 1.0$ and (c) $h_z = 1.4$. The color bar corresponds to the z-component, and the arrows indicate the planar components of the spins. (d)-(f)
The corresponding spin structure factor for the three values of the magnetic field $h_z$.}
\end{figure*}

\begin{figure*}
\centering
\includegraphics[width=1 \textwidth,angle=0,clip=true]{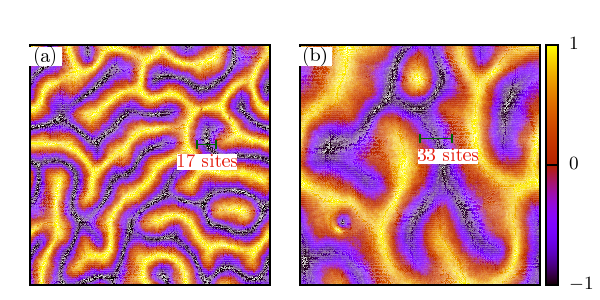}
\caption{ (Color online) Real space view of spin configurations in $N= 240^2$ system at (a) $T /t = 0.01$ and $\lambda/t = 0.1$ and (b) $T /t = 0.005$ and $\lambda/t = 0.05$  The color bar corresponds to the $z$ component, and the arrows indicate the planar components of the spins.}

\end{figure*}

\begin{figure*}
\centering
\includegraphics[width=1 \textwidth,angle=0,clip=true]{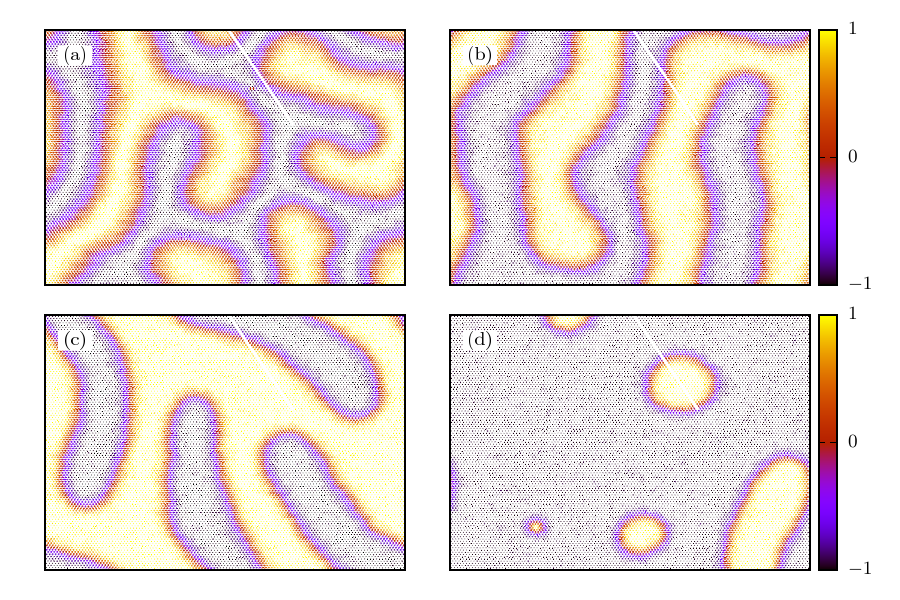}
\caption{ (Color online) Representative spin configurations at $T /t = 0.01$ and $\lambda/t = 0.1$ for (a) $A_u = 0.02$, (b) $A_u = 0.04$ and (c) $A_u = 0.06$, (d) $A_u = 0.08$. The color bar corresponds to the $z$ component, and the arrows indicate the planar components of the spins.}
\end{figure*}